\documentclass[RNAAS]{aastex7}

\renewcommand{\baselinestretch}{1.1}
\usepackage{textcomp}
\usepackage{gensymb}
\usepackage{epstopdf}
\usepackage{amsmath}
\usepackage{graphicx}
\usepackage{hyperref}
\usepackage{multirow}
\usepackage[nameinlink]{cleveref}
\usepackage[multiple]{footmisc}
\crefname{paragraph}{\S}{\S\S} 
\defcitealias{kluter:2025a}{Kl{\"u}ter \& Huston et al.}

\shorttitle{\texttt{THOR} and \texttt{HAMRR}} 
\shortauthors{Terry \& Anderson}

\begin{document}
\title{\textbf{\texttt{THOR} and \texttt{HAMRR}}}

\author[0000-0002-5029-3257]{Sean K. Terry}
\affiliation{Department of Astronomy, University of Maryland, College Park, MD 20742, USA}
\affiliation{Code 667, NASA Goddard Space Flight Center, Greenbelt, MD 20771, USA}
\email[show]{skterry@umd.edu}

\author[0000-0003-2861-3995]{Jay Anderson}
\affiliation{Space Telescope Science Institute, 3700 San Martin Drive, Baltimore, MD 21218, USA}
\email{jayander@stsci.edu}

\begin{abstract}

\small \noindent We present the Terry Hubble Observations of Roman (\texttt{THOR}) data reduction pipeline and Hubble Advanced Mining Routine for Roman (\texttt{HAMRR}). \texttt{THOR} is designed to reduce HST Wide-field Camera 3 (WFC3) and Advanced Camera for Surveys (ACS) imaging data taken as part of program GO-17776: \textit{A Precursor Survey of the Roman Galactic Bulge Time Domain Fields} \citep{terry:2024prop}. The primary function of \texttt{HAMRR} is to query the \texttt{THOR}-derived catalog via a typical cone search algorithm. Output products from \texttt{HAMRR} include calibrated photometry and astrometry for detected point sources in the HST catalog. The package supports additional output products such as cutout images, auto-generated color-magnitude diagrams, luminosity functions, and more. The \texttt{HAMRR} tool can be used in coordination with analyses of Roman, Rubin/LSST, and Euclid Galactic Bulge targets in the near future.
\\
\\
\textit{Subject headings}: Galactic bulge: Stellar populations \textemdash \, Astronomy software \textemdash \, HST \textemdash \, Roman\\
\end{abstract}


\section{Introduction} \label{sec:intro}
\indent One of the deepest observations ever conducted will be the Nancy Grace Roman Space Telescope's Galactic Bulge Time Domain Survey (GBTDS). In its primary survey area, the GBTDS will monitor five fields totaling approximately 1.4 square degrees using the Wide Field Instrument (WFI). A unique aspect of the GBTDS, relative to other bulge surveys, is its combination of large areal coverage and high angular resolution. In particular, Roman will resolve hundreds of millions of main-sequence stars \citep{penny:2019a}. Images at other wavelengths, or at times significantly before or after the survey, that also resolve and detect even a fraction of these stars, will provide additional information about the stellar populations toward the center of the Milky Way. \\
\indent In terms of angular resolution and sensitivity, HST is exceptionally well matched to Roman. Although HST's field of view is ${\sim} 100\times$ smaller than Roman's, the native pixel sizes for HST's cameras are $2-3 \times$ smaller than those of the Roman WFI. In addition, HST's imaging instruments provide access to bluer wavelengths that are complementary to Roman’s primarily near-infrared coverage. HST has previously been used to image the Galactic bulge along multiple sight lines to study populations \citep{Kozlowski:2006a, Clarkson:2008a}, ages \citep{Calamida:2014a, Baldwin:2016a}, and the star formation history \citep{bernard:2018a}. These prior campaigns have conducted pencil-beam observations of targeted bulge regions and therefore cover only a small fraction of the total bulge area. \cite{terry:2026a} recently conducted a wide area survey of this bulge region with HST utilizing both the Advanced Camera for Surveys (ACS) and Wide-field Camera 3 (WFC3) in parallel-imaging mode. This program has several science goals, one of which is to extend the time baseline between the earliest and latest epochs of space-based high resolution imaging for many millions of stars in these fields. The bluer HST coverage also expands the stellar spectral energy distribution (SED) range for characterization of exoplanet host stars, low-mass M dwarfs, flaring stars, oscillating red giants, and many others. Further descriptions of the science that is enabled by this survey can be found in Sec. 3 of \cite{terry:2026a}.\\
\indent While the scope of the HST bulge survey is broad, the project is largely designed to support the Roman GBTDS mission. This paper is an intermediate product between the published early results of \cite{terry:2026a} and the full data release of \cite{terry:inprep}. In this work we give an overview of the data reduction pipeline (DRP), resulting data products, and software tools which have been developed as part of the program.


\section{Data Reduction} \label{sec:drp}
\begin{figure}[!htb]
\includegraphics[width=\linewidth]{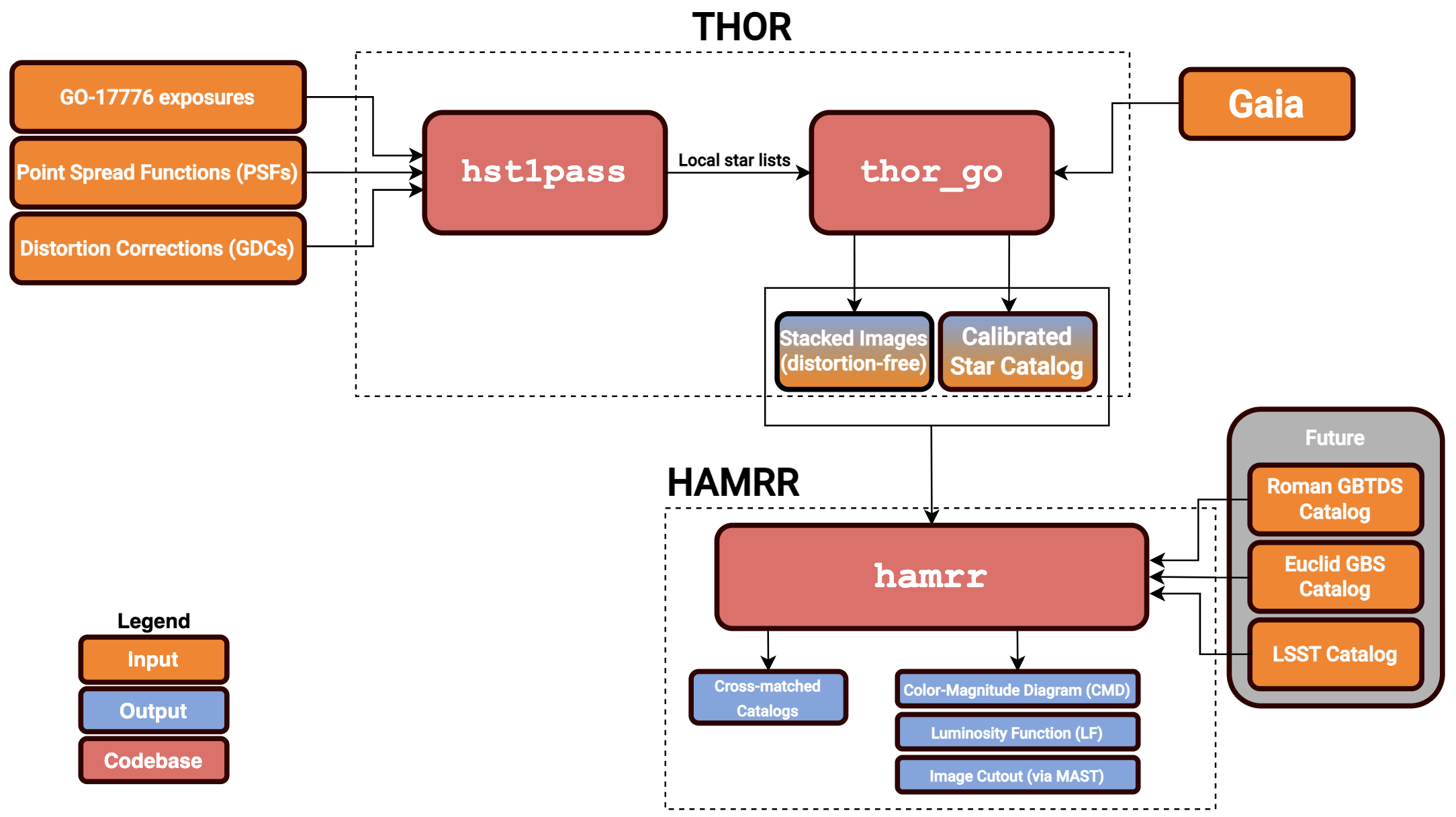}
\centering
\caption{A diagram depicting the core functionality of \texttt{THOR} and \texttt{HAMRR}. Orange boxes represent inputs, blue boxes represent outputs, and red boxes represent the codebases.} \label{fig:drp-diagram}
\end{figure}

Following standard procedures, we access the calibrated HST exposures in both ACS and WFC3 cameras from the Mikulski Archive for Space Telescopes (MAST)\footnote{\url{https://mast.stsci.edu/portal/Mashup/Clients/Mast/Portal.html}}. Reducing these data with \texttt{THOR} requires two distinct steps:
\begin{enumerate}
    \item \texttt{hst1pass} \citep{anderson:2006a, anderson:2022a}
    \item \texttt{thor\char`_go} (this work)
\end{enumerate}

\noindent where \texttt{hst1pass} is a widely used PSF-fitting algorithm, particularly well-suited for photometry and astrometry on crowded-field point sources such as the Galactic bulge. \texttt{thor\char`_go} is largely based on a separate routine from \cite{anderson:2022a} called \texttt{hst2collate}, however we have made significant modifications in order to handle the unique dither strategy and significant exposure time difference for the GO-17776 images \citep{terry:2024prop, terry:2026a}. In Figure \ref{fig:drp-diagram} we show a diagram of \texttt{THOR}, which takes various files as input and generates calibrated star catalogs and stacked reference images. We note several low-level and diagnostic files (e.g., perturbed PSF models, subtracted images, region fields, etc) are generated as outputs of both \texttt{hst1pass} and \texttt{thor\char`_go}. \\
\indent Source finding and collation produces catalogs with roughly $70,000 - 100,000$ stars per ACS field, and $40,000-70,000$ stars per WFC3 field. We establish the astrometric reference frame in \texttt{thor\char`_go} using well-measured Gaia DR3 sources \citep{vallenari:2023a}. Well-measured in this case means that we use only Gaia stars with RUWE $< 1.3$ and AENS $< 2$ mas in the transformation.

\subsection{Advanced Products} \label{subsec:data-products}
The first high-level science product (HLSP) published as part of this program is the ``early-release" catalog of \cite{terry:2026a}. It includes nearly 800,000 point sources measured in eight distinct bulge fields. We selected these fields to sample a wide sky area within the expected Roman GBTDS footprint. The selected fields also have significantly varying interstellar extinction and reddening (see Table 3 of \cite{terry:2026a}).\\
\indent At present, the data for nearly all 354 individual fields have been reduced via \texttt{THOR}. This ``intermediate" catalog is currently available via download on the \texttt{THOR} GitHub\footnote{\url{https://github.com/skterry/THOR}}. We stress that this intermediate catalog has yet to be peer-reviewed. As such it should be considered a ``shared-risk" data product until it officially becomes an HLSP \citep{terry:inprep}. The intermediate catalog has ${\sim}22$ million detected sources across 332 HST fields in the bulge. Standard photometry, astrometry, and various fitting metrics are included in all catalogs produced by \texttt{THOR}.\\
\indent We have developed additional software for querying the \texttt{THOR}-derived catalogs, the Hubble Advanced Mining Routine for Roman (\texttt{HAMRR}) tool. \texttt{HAMRR} is a modular Python routine that takes the stellar catalog as input, queries sky coordinates (via cone search), and returns a table of stars that meet the search criteria. Figure \ref{fig:drp-diagram} shows a diagram of \texttt{HAMRR}, where the outputs of \texttt{THOR} are given to \texttt{HAMRR} as input. Additionally, in the future, external catalogs (e.g., Roman, Euclid, Rubin) can be ingested by \texttt{HAMRR} as additional star lists to interact with. \texttt{HAMRR} can (optionally) return additional products such as luminosity functions, color-magnitude diagrams, and image cutouts centered on the queried sky coordinates.

\section{Results and Discussion} \label{sec:results}
In this research note we introduced \texttt{THOR} and \texttt{HAMRR}, two software routines designed to reduce, analyze, and interact with HST data taken as part of GO-17776 \citep{terry:2024prop}. We briefly described the early-release HLSP published in \cite{terry:2026a}, as well as a new, intermediate catalog available via GitHub. Data release 1 (DR1) from the HST bulge survey is currently being prepared and will be published in \cite{terry:inprep}.\\
\indent Lastly, the \texttt{HAMRR} query tool allows one to perform a cone search on the HST catalog, returning all detected sources within the search radius, along with optional output products and figures. This tool can be used in coordination with future external catalogs like Roman GBTDS, Euclid GBS, and Rubin/LSST to perform joint analyses of millions of sources in these highly-crowded galactic bulge fields. Properly interpreting these point source catalogs and performing reliable stellar characterization will be of extreme importance leading into the Roman era.

\noindent \textit{Software}: Astropy \citep{astropy:2018, astropy:2022}, hst1pass \citep{anderson:2022a}, Matplotlib \citep{hunter:2007a}, Numpy \citep{harris:2020array} THOR/HAMRR (this work).

\noindent \textit{Use of AI Tools}: \texttt{THOR} is entirely based on legacy \texttt{Fortran} code developed by J. Anderson and S. K. Terry. The \texttt{HAMRR} code was largely developed by S. K. Terry, with assistance from Claude (Anthropic). The authors declare that no genAI or AI-assisted
technologies were used in the preparation of this manuscript.

\bibliographystyle{aasjournal}
\bibliography{thor_hamrr.bib}

\end{document}